\documentclass[aps,prc,reprint,superscriptaddress,longbibliography,nofootinbib,floatfix]{revtex4-2}
\usepackage[T1]{fontenc}
\usepackage{amssymb}
\usepackage{amsmath}
\usepackage{graphicx}
\usepackage{booktabs}
\usepackage{xurl}
\usepackage{hyperref}
\hypersetup{pdftitle={Softer charged-hadron spectra in Ne+Ne than in O+O collisions at the same multiplicity at sqrt(sNN) = 5.36 TeV}, pdfauthor={Kishora Nayak}}

\newcommand{\pT}{p_{\mathrm{T}}}
\newcommand{\sNN}{\sqrt{s_{\mathrm{NN}}}}
\newcommand{\avg}[1]{\langle #1\rangle}
\newcommand{\dNdeta}{\mathrm{d}N_{\mathrm{ch}}/\mathrm{d}\eta}

\newcommand{\lamNO}{\lambda_{\mathrm{Ne/O}}}

\begin{document}

\title{Softer charged-hadron spectra in Ne+Ne than in O+O collisions at the same multiplicity at \texorpdfstring{$\sNN=5.36$~TeV}{sqrt(sNN) = 5.36 TeV}}

\author{Kishora Nayak}
\email{k.nayak1234@gmail.com}
\affiliation{Department of Physics, Panchayat College, Bargarh 768028, Odisha, India}

\date{\today}

\begin{abstract}
Charged-hadron spectra measured by the ATLAS Collaboration in O+O
and Ne+Ne collisions at $\sNN=5.36$~TeV are compared at the same
charged-particle multiplicity instead of at the same centrality,
to test whether the shape of the transverse-momentum spectrum at
a given multiplicity depends on the size of the colliding system.
 At equal multiplicity, Ne+Ne collisions have a
lower mean transverse momentum $\avg{\pT}$ than O+O collisions.
Without any spectral fit, the published $\avg{\pT}$ values give a
Ne+Ne/O+O ratio of 0.9822 at 3--4\% centrality, 6.6 standard deviations below unity even if the O+O systematic uncertainties are taken as uncorrelated
between centrality classes. At $0.4<\pT<2$~GeV the Ne+Ne spectrum is
described by the O+O spectrum with the transverse-momentum axis
compressed by a single stretch factor, $\lamNO=0.969$ in the 3--4\% class. It lies 9.5 or 3.3 standard deviations below unity, depending on whether the common part of the O+O systematic
uncertainties cancels between centrality classes. This ordering is
expected if Ne+Ne collisions have a larger transverse area at the same
multiplicity and a larger area gives a softer spectrum. The softening decreases towards
peripheral collisions and is the same at forward pseudorapidity. In
central collisions the spectral shape follows the centrality class
more closely than the multiplicity. The same procedure is tested in
Xe+Xe and Pb+Pb collisions.
\end{abstract}

\maketitle

\section{Introduction}
\label{sec:intro}
Collisions of light nuclei at the LHC extend the study of collective phenomena
to systems intermediate in size between those formed in proton--nucleus and
heavy-ion collisions~\cite{Brewer:2021kiv,Nagle:2018nvi,Grosse-Oetringhaus:2024bwr,Huang:2019tgz}.
In 2025, the LHC provided $^{16}$O+$^{16}$O and $^{20}$Ne+$^{20}$Ne collisions at
$\sNN=5.36$~TeV. The two nuclei differ by four nucleons
and in shape: $^{16}$O is expected to have a compact, $\alpha$-clustered structure
and $^{20}$Ne an elongated one~\cite{Broniowski:2013dia,Zhang:2024vkh,Giacalone:2024luz}.
The first measurements of anisotropic flow in O+O and Ne+Ne collisions indicate
that the final-state anisotropy is sensitive to the initial nuclear
geometry~\cite{ALICE:2025luc,ATLAS:2025nnt,CMS:2025tga}, as predicted by model
calculations~\cite{Lim:2018huo,Sievert:2019zjr,Schenke:2020mbo,Nijs:2021clz,Giacalone:2024luz}.

Anisotropic flow is sensitive mainly to the shape of the initial state,
whereas the transverse-momentum ($\pT$) spectrum is sensitive to its size.
At a fixed entropy per unit rapidity, approximately at a fixed $\dNdeta$, a smaller transverse
size gives a larger entropy density and a stronger pressure gradient, and
therefore a larger mean transverse momentum $\avg{\pT}$ and a harder
spectrum~\cite{Gardim:2019xjs,CMS:2024sgx,Giacalone:2021clp}. In a hydrodynamic
description the hardening comes from a stronger radial expansion of the
medium~\cite{Schnedermann:1993ws,Heinz:2013th}. The same relation between size
and $\avg{\pT}$ is used to interpret event-by-event $\avg{\pT}$ fluctuations as
fluctuations of the initial size~\cite{Broniowski:2009fm,Bozek:2012fw}, and the
$\pT$ dependence of these fluctuations has recently been measured in Pb+Pb
collisions~\cite{ATLAS:2025ztg,ALICE:2025iud}. A larger $\avg{\pT}$ for a
smaller size at the same multiplicity is also expected without a
hydrodynamic expansion: in initial-state saturation models, $\avg{\pT}$
scales with the saturation momentum, whose square is proportional to the
multiplicity per unit transverse area~\cite{McLerran:2013oju}.

A test of the size dependence of $\avg{\pT}$ with O+O and Ne+Ne collisions
requires a common reference for the two colliding systems. The same
centrality class does not give the same multiplicity in two different
systems. The ATLAS Collaboration has measured charged-particle spectra,
$\dNdeta$ and $\avg{\pT}$ in O+O and Ne+Ne collisions in fine centrality
classes~\cite{ATLAS:2026zgq}. At equal centrality, the Ne+Ne/O+O ratio of
the spectra is nearly independent of $\pT$ and the $\avg{\pT}$ values
agree within 0.3\% for $\pT>0.27$~GeV~\cite{ATLAS:2026zgq}, while the
Ne+Ne multiplicity is larger by a factor between 1.05 and 1.22, depending on
centrality~\cite{ATLAS:2026zgq}. A comparison at the same $\dNdeta$
removes the difference in multiplicity, as was done for $\avg{\pT}$ in
pp, p+Pb and Pb+Pb collisions~\cite{ALICE:2013rdo}, where the smaller
system has the higher $\avg{\pT}$ at the same multiplicity. Those systems
differ strongly in size and overlap in multiplicity only over a limited
range. The O+O and Ne+Ne nuclei are close in size, and both systems were
measured in the same detector in fine centrality classes over
overlapping multiplicity ranges, so that the comparison can be made
class by class, with systematic uncertainties that partly cancel in the
published Ne+Ne/O+O ratios. Equal
multiplicity, however, does not necessarily imply equal spectral
shapes: a recent study of the $\Lambda/K^0_{\mathrm{S}}$ ratio in O+O,
p+Pb and Pb+Pb collisions finds that its $\pT$ dependence is ordered
better by the transverse density, calculated in a Glauber model, than by
the multiplicity alone~\cite{Neeraj:2026jkk}, so that systems of
different size can have different spectral shapes at the same
multiplicity.

The aim of this work is to test whether the charged-hadron spectra of
Ne+Ne and O+O collisions have the same shape when the two systems are
compared at the same multiplicity. If the spectral shape were fixed by
the multiplicity alone, the two spectra would coincide; a difference at
the same $\dNdeta$ would show that the shape depends on more than the
multiplicity, for example on the size of the colliding system. Since
$^{20}$Ne is the larger nucleus, such a size dependence would make the
Ne+Ne spectrum softer. The difference is quantified by a stretch factor
$\lamNO$: the Ne+Ne spectrum is described by the O+O spectrum at the same
multiplicity with every $\pT$ value multiplied by $\lamNO$, so that
$\lamNO<1$ means a softer Ne+Ne spectrum. For charged hadrons $\lamNO$ is
an effective parameter, an average over the hadron mixture dominated by
pions. The ratio of the published $\avg{\pT}$
values at matched multiplicity needs neither a spectral fit nor the
rescaling hypothesis and is the most direct measure of the softening.
The spectra at forward pseudorapidity and a Xe+Xe/Pb+Pb control
comparison test the detector-related part of the analysis and whether
the multiplicity matching by itself produces a softening.

The paper is organised as follows. Section~\ref{sec:method} describes the
data, the multiplicity matching, the rescaling fit, the evaluation of the
uncertainties and the validation of the method. Section~\ref{sec:results}
presents the spectra at matched multiplicity, the multiplicity dependence
of $\lamNO$ at mid and forward pseudorapidity, the ratio of the published
$\avg{\pT}$ values at matched multiplicity, and the Xe+Xe/Pb+Pb control
comparison. Section~\ref{sec:discussion} relates the equal-multiplicity
comparison to the equal-centrality comparison, discusses the results in
terms of system size and, with a blast-wave parametrisation, of the
freeze-out parameters, and determines the multiplicity ratio at equal
$\avg{\pT}$. Section~\ref{sec:summary} gives the summary and outlook,
and  Appendix~\ref{app:matching} lists the multiplicity matching for all
centrality classes and the stretch factors for all compared classes.

\section{Data and analysis method}
\label{sec:method}
\subsection{Input data}
\label{sec:data}
The analysis uses the charged-hadron measurements in O+O and Ne+Ne
collisions at $\sNN=5.36$~TeV published by the ATLAS
Collaboration~\cite{ATLAS:2026zgq}, in centrality classes 1\% wide from
0 to 5\% and 5\% wide from 5 to 80\%. The 70--80\% class is the reference of the published ratios given below; the other wider classes (0--5\% and 10\% wide from 20 to 80\%) are used only to test the multiplicity assignment (Sec.~\ref{sec:unc}). The main inputs are the $\pT$
spectra of the two systems at $|\eta|<0.5$ for $0.21<\pT<5$~GeV and two
ratios of these spectra: each O+O centrality class to the O+O 70--80\%
class, and Ne+Ne to O+O at equal centrality. The multiplicity is taken
from $\dNdeta$ at $|\eta|<0.5$, extrapolated to $\pT=0$, and from the
Ne+Ne/O+O ratio of $\dNdeta$. The analysis is repeated at
$1.5<|\eta|<2.25$ with the spectra and ratios measured there, keeping the
mid-rapidity multiplicities. The ratio of $\avg{\pT}$ at matched multiplicity
(Sec.~\ref{sec:meanpt}) and the multiplicity ratio at equal $\avg{\pT}$
(Sec.~\ref{sec:ruler}) use $\avg{\pT}$ and $\dNdeta$ for
$\pT>0.27$~GeV and their Ne+Ne/O+O ratios. Each data point has a
statistical and a total systematic uncertainty; their treatment is
described in Sec.~\ref{sec:unc}. The control comparison
(Sec.~\ref{sec:xepb}) uses the charged-hadron spectra, $\avg{\pT}$ and
$\dNdeta$ measured by the ALICE Collaboration in Xe+Xe and Pb+Pb
collisions~\cite{ALICE:2018hza,ALICE:2018vuu} and the $\avg{[\pT]}$
measured by the ATLAS Collaboration~\cite{ATLAS:2024jvf}; the blast-wave
study (Sec.~\ref{sec:interp}) uses the ALICE $\pi$, K and p spectra in
Pb+Pb collisions~\cite{ALICE:2019hno}.

\subsection{Multiplicity matching}
\label{sec:matching}
To compare the two systems at equal multiplicity, the O+O spectrum is
needed at the $\dNdeta$ of each Ne+Ne centrality class $c$. The Ne+Ne
$\dNdeta$ is obtained from the O+O value of the same class and the
published ratio,
\begin{equation}
\dNdeta\big|_{\mathrm{Ne}}(c)=r_{\mathrm{Ne/O}}(c)\times\dNdeta\big|_{\mathrm{O}}(c),
\label{eq:match}
\end{equation}
where $r_{\mathrm{Ne/O}}(c)$ is the published Ne+Ne/O+O $\dNdeta$ ratio
and $\dNdeta|_{\mathrm{O}}(c)$ the O+O multiplicity of the same centrality
class. Equation~\eqref{eq:match} reproduces the published Ne+Ne multiplicity
within 0.01\%. The matched value is used instead of the published
Ne+Ne value because the matched O+O spectrum lies on the same O+O multiplicity
scale: an overall uncertainty of that scale cancels, and the matching depends mainly on the uncertainty of $r_{\mathrm{Ne/O}}$, 0.1\% in the most central
classes compared with 2.3--2.5\% for the Ne+Ne multiplicity itself.
The multiplicity is matched at $|\eta|<0.5$, where the spectra are
measured. The published Ne+Ne/O+O ratio of $\dNdeta$ changes by less
than 0.9\% between $|\eta|<0.5$ and any $|\eta|<2.4$ in all classes,
within the $\pm1\%$ variation of the matched multiplicity.

The O+O spectrum at the multiplicity $m=\dNdeta|_{\mathrm{Ne}}(c)$ is
obtained by interpolation between the two O+O classes 1 and 2 whose
multiplicities satisfy $m_1>m>m_2$. The interpolation is linear in the
logarithm of the spectrum and the logarithm of the multiplicity,
\begin{equation}
\ln f_{\mathrm{O}}(\pT;m)=w_1\ln f_{\mathrm{O},1}(\pT)+w_2\ln f_{\mathrm{O},2}(\pT),
\label{eq:interp}
\end{equation}
with $w_1=(\ln m-\ln m_2)/(\ln m_1-\ln m_2)$ and $w_2=1-w_1$,
applied bin by bin, where $f=(1/\pT)\,\mathrm{d}N/\mathrm{d}\pT$. A logarithmic interpolation is used because the yields in a given $\pT$ bin vary approximately as a power of the multiplicity, for which Eq.~\eqref{eq:interp} is exact; a quadratic interpolation over three classes is used as a systematic variation.

For example, the Ne+Ne 3--4\% class ($\dNdeta=145.0$) is matched by the O+O 0--1\% and 1--2\% classes with weights 0.79 and 0.21; the weights of all classes are listed in Appendix~\ref{app:matching}. The three most central Ne+Ne classes (0--3\%) have a higher multiplicity than the most central O+O class ($\dNdeta=148.4$) and cannot be matched, so the comparison starts at Ne+Ne 3--4\%. Each class is assigned a single multiplicity, although its spectrum averages over a range of multiplicities; this approximation is tested in Sec.~\ref{sec:unc}.

\subsection{Transverse-momentum rescaling}
\label{sec:rescaling}
The Ne+Ne spectrum is compared with the matched O+O spectrum under the hypothesis that the two spectra differ only by a stretch of the $\pT$ axis,
\begin{equation}
\left.\frac{\mathrm{d}N_{\mathrm{Ne}}}{\mathrm{d}\pT}\right|_{\pT}
=\frac{C}{\lamNO}\left.\frac{\mathrm{d}N_{\mathrm{O}}}{\mathrm{d}\pT}\right|_{\pT/\lamNO},
\label{eq:resc}
\end{equation}
where $C$ is a free normalisation, which does not enter the fit. Since the spectra are published as yields in $\pT$ bins, the fit uses the integrated form
\begin{equation}
N_{\mathrm{Ne}}(l,h)=C\,N_{\mathrm{O}}\!\left(\frac{l}{\lamNO},\frac{h}{\lamNO}\right),
\label{eq:resc_int}
\end{equation}
where $N(l,h)=\int_l^h(\mathrm{d}N/\mathrm{d}\pT)\,\mathrm{d}\pT$ is the yield in the interval $l<\pT<h$. Equation~\eqref{eq:resc_int} follows from Eq.~\eqref{eq:resc} by substituting $\pT\to\pT/\lamNO$ in the integral. For the invariant spectrum Eq.~\eqref{eq:resc} reads $f_{\mathrm{Ne}}(\pT)=C\,\lamNO^{-2}f_{\mathrm{O}}(\pT/\lamNO)$.

The rescaled intervals $[l/\lamNO,h/\lamNO]$ do not coincide with the published bins. The yields in the rescaled intervals are computed by modelling the published spectrum inside each bin as an exponential in $\pT$ with the published value at the bin centre and the local logarithmic slope. For each trial value of $\lamNO$ the logarithmic difference in bin $i$, with edges $l_i,h_i$, is
\begin{equation}
d_i=\ln N_{\mathrm{Ne}}(l_i,h_i)-\ln N_{\mathrm{O}}\!\left(\frac{l_i}{\lamNO},\frac{h_i}{\lamNO}\right)
\label{eq:di}
\end{equation}
and the fit minimises
\begin{equation}
\chi^2(\lamNO)=\sum_{i\in S}\frac{(d_i-\bar d)^2}{\sigma_i^2},\qquad \bar d=\frac{\sum_{i\in S}d_i/\sigma_i^2}{\sum_{i\in S}1/\sigma_i^2},
\label{eq:chi2}
\end{equation}
over the fit range $S$, the 20 published $\pT$ bins between 0.405 and 2.05~GeV (referred to as $0.4<\pT<2$~GeV). Subtracting the weighted mean $\bar d$ removes the constant $\ln C$ from $d_i$, so only the shapes of the spectra enter the fit. The point-to-point uncertainty $\sigma_i$ combines the published statistical and systematic uncertainties of the spectra and of their ratios, as described in Sec.~\ref{sec:unc}. The $\sigma_i$ are treated as uncorrelated between $\pT$ bins; the part of the systematic uncertainties that is correlated in $\pT$ is accounted for by the coherent variations of the spectra described in Sec.~\ref{sec:unc}.

Since the hadron masses matter most at low $\pT$, a single stretch factor describes the mixture of hadrons least well there: in the most central classes the $\chi^2$ of the fit grows faster than the number of degrees of freedom when the fit starts below about 0.35~GeV. The fit therefore starts at 0.405~GeV, although the spectra are published from 0.21~GeV, and ends at 2.05~GeV to restrict it to the soft part of the spectra; for Ne+Ne 3--4\% it gives $\chi^2/\mathrm{NDF}=5.9/18$.

The departure from pure rescaling in each bin is the residual of Eq.~\eqref{eq:chi2} at the fitted $\lamNO$,
\begin{equation}
D_i=d_i-\bar d
\label{eq:D}
\end{equation}
evaluated for $0.405<\pT<4.61$~GeV with $\bar d$ taken from the fit range; the highest published bin, 4.61--5~GeV, is excluded because for $\lamNO<1$ its rescaled interval lies beyond the measured O+O range. By construction $D_i$ vanishes on average in $S$, so a departure from pure rescaling outside $S$ appears as $D_i\neq0$.

At $\lamNO=1$ the same quantity is the logarithmic ratio of the Ne+Ne spectrum to the O+O spectrum at the matched multiplicity, shown in Fig.~\ref{fig:spectra}(a),
\begin{equation}
\ln R_{\mathrm{matched},i}=\ln\frac{N_{\mathrm{Ne}}(l_i,h_i)}{N_{\mathrm{O}}(l_i,h_i)}-\left\langle\ln\frac{N_{\mathrm{Ne}}}{N_{\mathrm{O}}}\right\rangle_{S},
\label{eq:lnR}
\end{equation}
where $N_{\mathrm{O}}$ is the O+O yield at the matched multiplicity, Eq.~\eqref{eq:interp}, and $\langle\,\rangle_S$ is the weighted mean over the fit range with the weights $1/\sigma_i^2$ of Eq.~\eqref{eq:chi2}. The difference $\ln R_{\mathrm{matched},i}-D_i$ is then the ratio expected for an exact rescaling with the fitted $\lamNO$, drawn as curves in Fig.~\ref{fig:spectra}(a).

\subsection{Uncertainties and validation}
\label{sec:unc}
The point-to-point uncertainty $\sigma_i$ of Eq.~\eqref{eq:chi2} has a statistical and a systematic part. The statistical part is that of the Ne+Ne spectrum of class $c$ and of the O+O spectrum at the matched multiplicity $m$, added in quadrature; the O+O class-$c$ and 70--80\% spectra, which enter the two published ratios, do not contribute, since the published statistical uncertainties of the ratios are the quadrature sums of those of the spectra.

The systematic part combines the uncertainty of the Ne+Ne/O+O ratio with that of the O+O step from class $c$ to $m$, computed from the systematic uncertainties $\sigma_{\mathrm{O}}$ of the O+O ratios to the 70--80\% class. Where the published systematic uncertainties of the spectra and their ratios are asymmetric, the mean of the upward and downward values is used. In the fit range $\sigma_{\mathrm{O}}$ is below 1\% up to 1~GeV and rises to about 2\% at 2~GeV, whereas the Ne+Ne/O+O ratio at equal centrality is known to 0.1--0.3\% in central classes, so the O+O step dominates unless the correlation of $\sigma_{\mathrm{O}}$ between classes is used.

The published $\sigma_{\mathrm{O}}$ is nearly the same in all classes from 0--1\% to 35--40\% (0.76--0.86\% at 1~GeV), and its sources are common to all classes~\cite{ATLAS:2026zgq}. The part common to all classes cancels in the step, leaving the difference $|B(c)-B(m)|$ of the class-dependent parts $B$. If these change in the same direction for all sources between neighbouring classes, the difference is bounded from above by the quadrature difference $|\sigma_{\mathrm{O}}^2(c)-\sigma_{\mathrm{O}}^2(m)|^{1/2}$, which is used in all classes. This treatment rests on two assumptions that the published tables, which give no breakdown of the systematic uncertainty by source and no correlation between classes, cannot test: that the common part is the same in all classes and fully correlated, and that the class-dependent parts of all sources change in the same direction. All results are therefore also given in the uncorrelated limit defined below, which makes neither assumption. Above 40\% centrality the step is added linearly to the uncertainty of the Ne+Ne/O+O ratio, because the centrality determination (uncertain by $\pm2.2\%$ in O+O and $\pm2.5\%$ in Ne+Ne collisions~\cite{ATLAS:2026zgq}) affects both and does not cancel in that ratio.

The uncertainty built from these statistical and systematic parts is referred to as the \emph{nominal} point-to-point uncertainty. In the \emph{uncorrelated limit}, the conservative alternative, the published total uncertainties of the O+O class $c$ and of the matched O+O spectrum are added in quadrature. In both cases nine sources are evaluated by refitting $\lamNO$ and added in quadrature: (i) coherent distortions of the Ne+Ne/O+O ratio within its uncertainty (the largest of a tilt over 0.3--3~GeV, a tilt over the full range and a curvature); (ii) the same for the O+O step from $c$ to $m$; (iii) quadratic instead of linear interpolation; (iv) the matched multiplicity varied by $\pm1\%$, 10--14 times the published uncertainty of $r_{\mathrm{Ne/O}}$ in central classes, or by that uncertainty where it is larger (up to 2.2\% in peripheral classes); (v) matching with the fiducial $\dNdeta$ ($\pT>0.27$~GeV) instead of the extrapolated one; (vi) the fit range, the largest shift over the twelve combinations of lower edges 0.405, 0.5 and 0.6~GeV and upper edges 1.16, 1.48, 1.89 and 2.41~GeV; (vii) the low edge of the fit, the larger shift when the fit starts at 0.27 or 0.30~GeV; (viii) the uncertainty of the fit ($\Delta\chi^2=1$ with the nominal weights, which overestimate it; Sec.~\ref{sec:spectra}); and (ix) the fit weights, the larger shift with uncorrelated-limit or uniform weights. Sources (i), (ii) and (viii) use the same published point uncertainties, so their quadrature sum is conservative. The resulting total uncertainty of $\lamNO$ is denoted $\sigma_{\mathrm{nom}}$, or $\sigma_{\mathrm{unc}}$ in the uncorrelated limit. For the Ne+Ne 3--4\% class $\sigma_{\mathrm{nom}}$ is 0.0032, mainly from the variation of the matched multiplicity, the fit range, the low edge of the fit and the matching variable (0.0013--0.0016 each); $\sigma_{\mathrm{unc}}$ is 0.0094, dominated by the shape of the O+O step.

A closure test on an exactly rescaled published spectrum recovers $\lambda$ to $10^{-5}$, and an independent implementation, with a cubic-spline model of the spectra, its own multiplicity interpolation and an unweighted fit over $0.405<\pT<1.5$~GeV, agrees with the main analysis within $|\Delta\lambda|\le0.0011$ in the six classes tested.

The single multiplicity assigned to each class (Sec.~\ref{sec:matching}) is tested with the O+O data alone: a published wide class compared with narrower classes interpolated to the same multiplicity should give $\lambda=1$. The 10\%-wide classes from 20 to 80\% agree with unity within 0.001, and 0--5\% compared with the 1\%-wide classes deviates by about 0.002. Classes of different width enter the matching only for two Ne+Ne classes. Ne+Ne 5--10\% is matched by 1\%-wide O+O classes; since it spans a narrower multiplicity range than O+O 0--5\%, the effect is below 0.002, and it raises $\lamNO$. Ne+Ne 10--15\% is matched by the O+O 4--5\% and 5--10\% classes (Appendix~\ref{app:matching}), and the corresponding mixed-width O+O test differs from the tests between classes of equal width by about 0.001.

\section{Results}
\label{sec:results}
\subsection{Spectra at matched multiplicity}
\label{sec:spectra}
\begin{figure}[!t]
\centering
\includegraphics[width=\columnwidth]{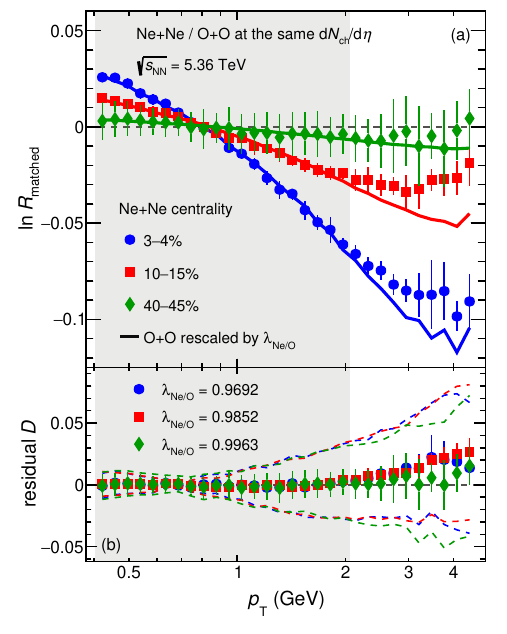}
\caption{(a) Logarithmic ratio $\ln R_{\mathrm{matched}}$ of the Ne+Ne
spectrum to the O+O spectrum at the same $\dNdeta$ in $|\eta|<0.5$ (data
from Ref.~\cite{ATLAS:2026zgq}). The curves show $\ln R_{\mathrm{matched}}-D$,
the ratio expected for an exact rescaling with the fitted $\lamNO$,
Eq.~\eqref{eq:resc}. (b) Residual $D$ from pure rescaling, Eq.~\eqref{eq:D}.
Bars show the nominal point-to-point uncertainties, statistical and
systematic combined (Sec.~\ref{sec:unc}); the dashed lines in (b) show the
uncertainty in the uncorrelated limit. The shaded area marks the fit range.}
\label{fig:spectra}
\end{figure}

Figure~\ref{fig:spectra}(a) shows $\ln R_{\mathrm{matched}}$,
Eq.~\eqref{eq:lnR}, for three Ne+Ne centrality classes. For the 3--4\%
centrality the ratio decreases steadily with $\pT$: at the same
multiplicity, Ne+Ne collisions produce relatively more particles at
low $\pT$ and fewer at high $\pT$ than O+O collisions. The curves
show the ratio expected for an exact rescaling with the fitted $\lamNO$.
A single stretch of the $\pT$ axis absorbs the difference in the soft
region: the residual $D$, shown in Fig.~\ref{fig:spectra}(b), stays
below 0.5\% in absolute value over the whole fit range in the three classes shown.
With the nominal uncertainties the fit gives $\chi^2/\mathrm{NDF}=0.005$--0.38
in all compared classes, because these uncertainties are dominated by
systematic parts correlated in $\pT$; they therefore do not test the
rescaling hypothesis. With the statistical uncertainties alone, $\chi^2$
is 24.2 for 18 degrees of freedom in the 3--4\% class and 15.2--34.9
in most other classes, but 68.1 and 49.0 in the 10--15\% and 15--20\%
classes; $\lamNO$ changes by at most 0.0005 with these weights. An
additional exponential tilt (a term linear in $\pT$ in $d_i$) lowers the
nominal $\chi^2$ by at most 4.2 but moves $\lamNO$ further from unity, by
0.0082 for 3--4\% and by up to 0.0156 in the other classes up to 70\%
centrality: over the fit range a stretch and a tilt are nearly degenerate,
and $\lamNO$ is the stretch of the one-parameter description.
Above the fit range, for $2.05<\pT<4.61$~GeV, the residual of the 3--4\%
and 10--15\% centrality classes is positive in every bin and grows with $\pT$,
reaching up to 2.6\% at 4.4~GeV, which is significant with the nominal
uncertainties only. For the 40--45\% centrality the residual stays within
the nominal uncertainty.
The stretch factors below are all fitted in the soft region.

\subsection{Multiplicity dependence of the stretch factor}
\label{sec:lambda}
\begin{figure}[!t]
\centering
\includegraphics[width=\columnwidth]{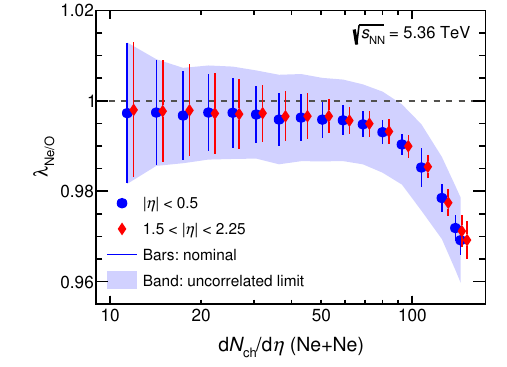}
\caption{Stretch factor $\lamNO$ as a function of the $\dNdeta$ of
the Ne+Ne centrality class, for the spectra at mid and forward
pseudorapidity. The bars show the nominal uncertainty and the band,
drawn for $|\eta|<0.5$ only, the uncertainty in the uncorrelated
limit. The forward points are shifted horizontally for clarity.}
\label{fig:lambda}
\end{figure}

Figure~\ref{fig:lambda} shows $\lamNO$ for different Ne+Ne centrality
classes from 3 to 80\%, plotted at their $\dNdeta$; the values are listed
in Appendix~\ref{app:matching}. In the most central class compared, 3--4\%,
$\lamNO=0.9692$: at $\dNdeta=145.0$
the Ne+Ne spectrum is the O+O spectrum compressed by 3.1\% along the $\pT$
axis. With decreasing multiplicity the stretch factor rises, to 0.9852 at
$\dNdeta=107.5$ (10--15\%) and 0.9931 at $\dNdeta=80.0$ (20--25\%). From
35\% centrality onwards $\lamNO$ is consistent with unity. With the nominal
uncertainties the deviation from unity is between $3.5\sigma$ (10--15\%)
and $9.5\sigma$ (3--4\%) in the classes from 3 to 20\% centrality, and below
$3\sigma$ between 20 and 35\%. From 40\% centrality onwards all values
lie between 0.9958 and 0.9974 at mid-rapidity, a common offset of about
0.3\% that is well within the uncertainties, which are correlated between
these classes.
In the uncorrelated limit the deviation is $3.3\sigma$
and $3.0\sigma$ for the 3--4\% and 4--5\% centrality classes and smaller
elsewhere: only the two classes matched to the O+O 0--2\% classes reach
$3\sigma$ in this limit, and the interpretation of these classes is
limited (Sec.~\ref{sec:eqcent}). Neighbouring values of $\lamNO$ are correlated, since
they share O+O input classes and the systematic uncertainties of the
published ratios, so the smooth trend in Fig.~\ref{fig:lambda} does not
by itself test the size of the uncertainties.

The spectra at $1.5<|\eta|<2.25$ come from the same events but from
different tracks, reconstructed in a different part of the detector.
The analysis was repeated with these spectra, for the same centrality
classes and matched multiplicities. The results, also listed in
Appendix~\ref{app:matching}, agree with those at mid-rapidity within
0.12\% in all compared centrality classes; for the 3--4\% class,
$\lamNO=0.9692\pm0.0042$. Since the centrality classes, the multiplicity
matching and the O+O interpolation are the same, this comparison tests
the track reconstruction and the fit, not the multiplicity matching.

\subsection{Mean transverse momenta at matched multiplicity}
\label{sec:meanpt}
\begin{figure}[!t]
\centering
\includegraphics[width=\columnwidth]{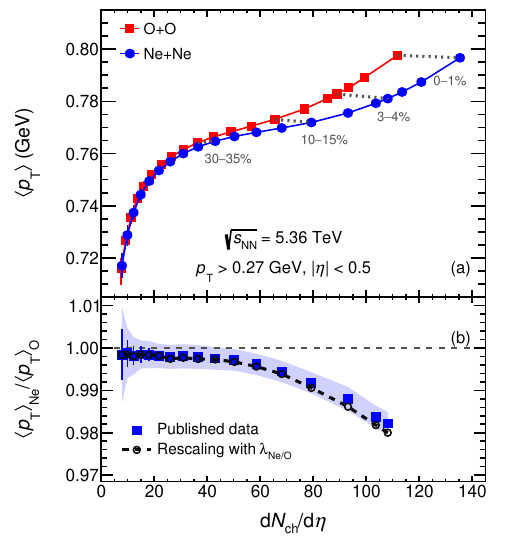}
\caption{(a) Published $\avg{\pT}$ versus $\dNdeta$ ($\pT>0.27$~GeV,
$|\eta|<0.5$) in O+O and Ne+Ne collisions~\cite{ATLAS:2026zgq};
solid lines join neighbouring classes and dotted lines classes of
the same centrality. (b) Ratio of the Ne+Ne and O+O $\avg{\pT}$ at
the same multiplicity, Eq.~\eqref{eq:r}, at the $\dNdeta$ of the
Ne+Ne class: published data (squares; bars: nominal uncertainty;
band: uncorrelated limit, larger side) and prediction from the
fitted $\lamNO$, Eq.~\eqref{eq:rpred} (open circles, dashed line).}
\label{fig:meanpt}
\end{figure}

The published $\avg{\pT}$ values measure the softening without the rescaling
hypothesis and without a spectral fit. They are derived from the same spectra
and provide a cross-check of $\lamNO$ that relies on fewer assumptions. Figure~\ref{fig:meanpt}(a) shows
the published $\avg{\pT}$ for $\pT>0.27$~GeV as a function of the published
$\dNdeta$ in the same $\pT$ range. The Ne+Ne values lie below the O+O
values at the same multiplicity, and classes of the same centrality have
nearly the same $\avg{\pT}$.

The ratio of $\avg{\pT}$ at matched multiplicity is built from the
published equal-centrality ratio and the O+O change from class $c$ to
the matched multiplicity,
\begin{equation}
\rho=\left[\frac{\avg{\pT}_{\mathrm{Ne}}}{\avg{\pT}_{\mathrm{O}}}
\right]_{c}\times\frac{\avg{\pT}_{\mathrm{O}}(c)}{\avg{\pT}_{\mathrm{O}}(m)},
\label{eq:r}
\end{equation}
with $\avg{\pT}_{\mathrm{O}}(m)$ interpolated linearly in $\ln\dNdeta$.
The ratio $\rho$ uses only the published tables: the $\avg{\pT}$ values,
and the $\dNdeta$ values to locate the matched multiplicity $m$.

For an exact rescaling, $\avg{\pT}$ in correspondingly scaled windows
satisfies $\avg{\pT}_{\mathrm{Ne}}=\lamNO\avg{\pT}_{\mathrm{O}}$, but the
published values are integrated above the same fixed threshold and include
the regions 0.27--0.405~GeV and above 2~GeV, which the fit does not
constrain. Figure~\ref{fig:meanpt}(b) and Table~\ref{tab:meanpt} therefore
compare $\rho$ with the prediction of the fitted $\lamNO$ in the same window,
\begin{equation}
\rho_{\mathrm{pred}}=\lamNO\,\frac{\avg{\pT}_{\mathrm{O}}[0.27/\lamNO,
\,5/\lamNO]}{\avg{\pT}_{\mathrm{O}}[0.27,\,5]},
\label{eq:rpred}
\end{equation}
where $\avg{\pT}_{\mathrm{O}}[l,h]$ is the mean $\pT$ between $l$ and $h$
(in GeV) of the O+O spectrum at the matched multiplicity, Eq.~\eqref{eq:interp},
continued above 5~GeV with the exponential shape of the last bin. The
published ratio lies closer to unity than the prediction, by about 0.002 in
the most central classes and by less towards peripheral collisions,
comparable to the uncertainty of $\lamNO$ propagated to the ratio; part of
the difference comes from matching on the fiducial $\dNdeta$, and its sign
is that expected from the positive residual $D$ above 2~GeV.

The published $\avg{\pT}$ values also give the significance of the
softening independently of the spectral uncertainties. The uncertainty
of $\rho$ is built as for the spectra (Sec.~\ref{sec:unc}), from the
published uncertainties of $\avg{\pT}$, but with the larger side of the
asymmetric O+O systematic uncertainty and a $\pm1\%$ variation of the
matched multiplicity, which dominates the nominal uncertainty in the
central classes. The deviation of $\rho$ from unity exceeds $4\sigma$ in all
classes up to 35--40\% with the nominal uncertainty and $3\sigma$ up to
the 10--15\% class in the uncorrelated limit (Table~\ref{tab:meanpt}).
In the uncorrelated limit the softening at 10--15\% centrality, where the
matched O+O classes are not among the most central, is thus significant
in $\rho$ ($3.3\sigma$) but not in $\lamNO$ ($1.5\sigma$), because the
$\avg{\pT}$ ratios are less sensitive than the spectral shape to the
$\pT$-dependent systematic uncertainties.

\begin{table}[!t]
\centering\footnotesize
\caption{Ratio $\rho$ of the Ne+Ne and O+O $\avg{\pT}$ (${\pT>0.27}$~GeV)
at matched multiplicity from the published tables, Eq.~\eqref{eq:r},
with the multiplicity matched on the fiducial $\dNdeta$, and the
rescaling prediction $\rho_{\mathrm{pred}}$ with the fitted $\lamNO$,
Eq.~\eqref{eq:rpred}, for selected Ne+Ne classes (all classes in
Fig.~\ref{fig:meanpt}). The last two columns give the deviation of
$\rho$ from unity in units of its nominal uncertainty $\sigma^{\rho}_{\mathrm{nom}}$
and of its uncertainty in the uncorrelated limit $\sigma^{\rho}_{\mathrm{unc}}$, both with the
larger side of the asymmetric O+O systematic uncertainty.}
\label{tab:meanpt}
\begin{tabular*}{\columnwidth}{@{\extracolsep{\fill}}ccccc}
\toprule
Ne+Ne class & $\rho$ & $\rho_{\mathrm{pred}}$ & $(1-\rho)/\sigma^{\rho}_{\mathrm{nom}}$
& $(1-\rho)/\sigma^{\rho}_{\mathrm{unc}}$\\
\midrule
3--4\% & 0.9822 & 0.9800 & 18.6 & 6.6\\
4--5\% & 0.9837 & 0.9818 & 16.7 & 6.1\\
5--10\% & 0.9880 & 0.9862 & 15.4 & 4.3\\
10--15\% & 0.9918 & 0.9906 & 13.3 & 3.3\\
15--20\% & 0.9945 & 0.9939 & 12.4 & 2.2\\
20--25\% & 0.9961 & 0.9956 & 10.1 & 1.5\\
30--35\% & 0.9975 & 0.9973 & 5.6 & 0.9\\
40--45\% & 0.9980 & 0.9977 & 3.0 & 0.7\\
\bottomrule
\end{tabular*}
\end{table}

\subsection{Control comparison: Xe+Xe versus Pb+Pb}
\label{sec:xepb}
Both the rescaling fit of the spectra (Sec.~\ref{sec:rescaling}) and the
ratio of $\avg{\pT}$ at matched multiplicity (Sec.~\ref{sec:meanpt}) rely on the same interpolation
in multiplicity, which could produce $\lambda<1$ for reasons unrelated to
geometry. To test whether the method itself produces $\lambda<1$, the same
procedure was applied to a pair in which the \emph{smaller} nucleus comes
first: Xe+Xe ($A=129$) at $\sNN=5.44$~TeV versus Pb+Pb ($A=208$) at
$\sNN=5.02$~TeV, using the ALICE charged-hadron spectra at $|\eta|<0.8$
in nine identical centrality classes and the ALICE
$\dNdeta$~\cite{ALICE:2018hza,ALICE:2018vuu}. If the geometric interpretation
is correct, the expected result is $\lambda\ge1$, opposite in sign to
Ne+Ne/O+O.

The two systems were measured at different energies, which requires a
correction. The energy dependence at fixed multiplicity was determined with
the same method from Pb+Pb collisions at 5.02 and 2.76~TeV, which give
$\lambda=1.052$--1.062, and was scaled linearly in $\ln\sNN$ to the interval
from 5.02 to 5.44~TeV (a factor 0.1343). This gives a factor $E\approx1.007$
for $\lambda$; the same procedure in the $\pT$ ranges of the $\avg{\pT}$
measurements gives 1.004--1.006. The Xe+Xe/Pb+Pb values are divided by
these factors. Minimum-bias pp spectra (measured at 5.02~TeV, interpolated
by ALICE to 5.44~TeV) give smaller factors (1.0031 for $\lambda$);
the difference is included as a systematic uncertainty. ALICE publishes a
single systematic uncertainty per point, so the correlated and uncorrelated
limits are defined by full cancellation and by quadrature addition of the
Xe+Xe and Pb+Pb systematic uncertainties.

\begin{figure*}[!t]
\centering
\includegraphics[width=\textwidth]{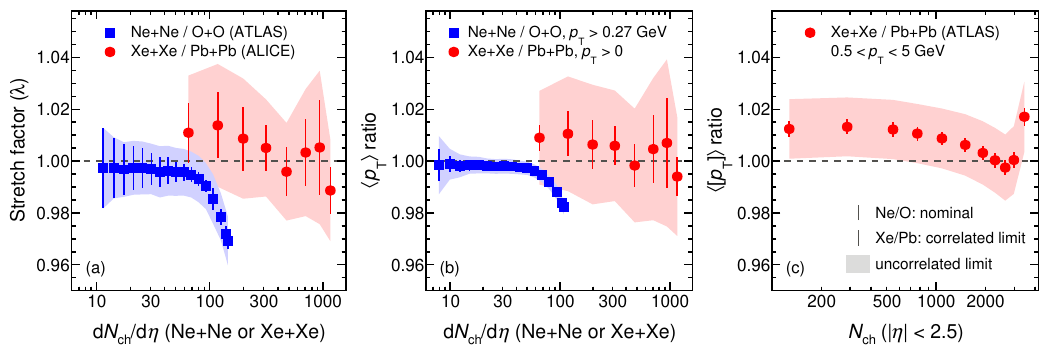}
\caption{Matched-multiplicity comparison for Ne+Ne/O+O~\cite{ATLAS:2026zgq},
larger nucleus first, and for the Xe+Xe/Pb+Pb control, smaller nucleus first.
(a) Stretch factor $\lambda$; Xe+Xe/Pb+Pb from ALICE~\cite{ALICE:2018hza,ALICE:2018vuu}.
(b) Ratio of the published $\avg{\pT}$ at matched multiplicity, Eq.~\eqref{eq:r};
Xe+Xe/Pb+Pb from ALICE values extrapolated to $\pT=0$. (c) Xe+Xe/Pb+Pb ratio of
$\avg{[\pT]}$ at the same $N_{\mathrm{ch}}$~\cite{ATLAS:2024jvf}. The Xe/Pb
ratios are corrected for the energy difference. Bars: nominal uncertainty
(Ne/O) or correlated limit (Xe/Pb); bands: uncorrelated limit.}
\label{fig:xepb}
\end{figure*}

After the energy correction, $\lambda(\mathrm{Xe/Pb})$ agrees with unity
within the uncertainties in all centrality classes that can be matched (the
70--80\% Xe+Xe class lies below the Pb+Pb multiplicity range), and no class
lies significantly below unity (Fig.~\ref{fig:xepb}(a)). The ALICE $\avg{\pT}$
tables, used as in Eq.~\eqref{eq:r}, give the same result (Fig.~\ref{fig:xepb}(b));
these values include an extrapolation to $\pT=0$, unlike the
fiducial ATLAS values used for Ne+Ne/O+O.
The event-averaged $\avg{[\pT]}$ for $0.5<\pT<5$~GeV and $|\eta|<2.5$,
measured by the ATLAS Collaboration in both systems as a function of the
number of charged particles $N_{\mathrm{ch}}$ in the same
acceptance~\cite{ATLAS:2024jvf}, gives a Xe+Xe/Pb+Pb ratio about 1\% above unity
up to $N_{\mathrm{ch}}\approx1500$, significant in the correlated limit only,
close to unity at higher $N_{\mathrm{ch}}$, and about 2\% above unity in the
ultra-central tail of Xe+Xe collisions (Fig.~\ref{fig:xepb}(c)).

The matched-multiplicity procedure therefore does not by itself produce a
softening as large as that observed in Ne+Ne/O+O collisions. The control
tests the method, not the geometric interpretation: the matched Pb+Pb
classes are not ultra-central, unlike the O+O classes matched to the most
significant Ne+Ne values, and the $\lambda(\mathrm{Xe/Pb})$ expected for
the much larger area difference of Xe+Xe and Pb+Pb is not evaluated here.
The ATLAS $\avg{[\pT]}$ ratios tend to lie slightly above unity, the sign
expected from geometry; the ALICE data alone do not establish this sign,
because their systematic uncertainties are large and the energy correction
is as large as the effect.

\section{Discussion}
\label{sec:discussion}
\subsection{Equal centrality versus equal multiplicity}
\label{sec:eqcent}
At equal centrality the Ne+Ne and O+O spectra have nearly the same
shape~\cite{ATLAS:2026zgq}, whereas at equal multiplicity the Ne+Ne spectrum
is softer. A decomposition of the fit relates the two observations. For the
3--4\% class, fitting the Ne+Ne spectrum to the O+O spectrum of the
\emph{same} centrality class gives $\lambda=0.9979$, a shape difference of
0.2\%, and fitting the O+O spectrum of that class to the O+O spectrum at the
matched multiplicity, which is 21\% higher, gives $\lambda=0.9712$. The
product, 0.9691, reproduces the matched value 0.9692. The softening at equal
multiplicity is therefore the hardening of the O+O spectrum when its
multiplicity is raised to the Ne+Ne value. The result combines two
published observations, the nearly equal spectral shapes at equal
centrality and the multiplicity dependence of the O+O spectrum, and
quantifies what follows from them: in central collisions the spectral
shape follows the centrality class more closely than the multiplicity.

The hardening is large because $\avg{\pT}$ rises steeply with multiplicity
in the most central collisions: the logarithmic slope
$a=\mathrm{d}\ln\avg{\pT}/\mathrm{d}\ln(\dNdeta)$ between the two most central
classes is 0.091 in O+O and 0.103 in Ne+Ne collisions; in O+O collisions it
falls to about 0.05 between 3 and 10\% centrality and to 0.018--0.023
between 10 and 35\% centrality. Such a steep rise is expected
when the multiplicity increases at a nearly fixed overlap geometry, so that
the density increases~\cite{Gardim:2019xjs,CMS:2024sgx,Samanta:2023amp}. The O+O collisions
matched to Ne+Ne 3--5\% are the 0--2\% most central ones, which have an
unusually high multiplicity for the O+O system size and whose selection
depends on the centrality estimator; this limits the interpretation of the
most central values of $\lamNO$. The softening persists where the matched
O+O classes are not among the most central: Ne+Ne 10--15\%, matched to O+O
4--10\%, gives a 1.5\% compression.

\subsection{Interpretation in terms of system size}
\label{sec:interp}
At the same $\dNdeta$ the two systems produce approximately the same
entropy per unit rapidity, so a larger transverse area of the initial
state in Ne+Ne than in O+O collisions would mean a smaller initial
entropy density. In a hydrodynamic expansion a lower entropy density
implies a lower effective temperature and weaker transverse pressure
gradients, and therefore a softer final-state
spectrum~\cite{Gardim:2019xjs,Broniowski:2009fm,Bozek:2012fw}.
The data agree qualitatively with this picture in two respects: the
ordering $\lamNO<1$, and the decrease of the effect towards peripheral
collisions, where the overlap region is set more by the positions of
a few participating nucleons than by the size of the nuclei.

The ordering does not by itself identify a hydrodynamic origin. In
initial-state saturation models, $\avg{\pT}$ is proportional to the
saturation momentum $Q_s$, with $Q_s^2$ proportional to
$(\dNdeta)/S_\perp$, where $S_\perp$ is the transverse area, and a larger
area at the same multiplicity also gives a softer
spectrum~\cite{McLerran:2013oju}. The data constrain the dependence of the
spectral shape on the transverse density, not the mechanism that produces
it. The hydrodynamic and EPOS calculations compared with the data in
Ref.~\cite{ATLAS:2026zgq} give the Ne+Ne/O+O ratios at equal centrality for
several models of the nuclear structure; whether they also reproduce
$\lamNO$ and $\rho$ at matched multiplicity, and whether these quantities
separate the nuclear-structure models, cannot be tested with published
material, since the model calculations are not part of the HEPData record
of the measurement.

The present analysis does not measure the entropy density, the
transverse size or the radial flow velocity directly. A stretch
of the $\pT$ axis can result from a change of the radial flow
velocity, of the kinetic freeze-out temperature or of both. For
charged hadrons, $\lamNO$ is an effective parameter of the particle
mixture, and the analysis assumes that the hadron composition is the same
in Ne+Ne and O+O collisions at the same multiplicity. This is supported by
the observation that ratios of identified-hadron yields in pp, p+Pb and
Pb+Pb collisions depend mainly on the multiplicity and little on the
colliding system~\cite{ALICE:2016fzo}. Identified-particle
spectra in centrality classes are not yet available for O+O and Ne+Ne
collisions, and an enhanced baryon-to-meson ratio, measured with strange
hadrons, has been observed in minimum-bias O+O relative to pp
collisions~\cite{CMS:2026aie}.
Matching the multiplicity also does not remove all differences in
the initial conditions, such as the shape and the fluctuations of
the initial density.

To relate $\lamNO$ to the flow velocity and the freeze-out temperature, a Boltzmann--Gibbs blast-wave
parametrisation~\cite{Schnedermann:1993ws} was fitted to the
identified-hadron spectra measured by the ALICE Collaboration in Pb+Pb
collisions at $\sNN=5.02$~TeV in the 50--60\% and 60--70\% centrality
classes~\cite{ALICE:2019hno}, whose multiplicities bracket those of the
compared central Ne+Ne classes. In the charged-hadron spectrum built from
these fits, either the mean transverse flow velocity $\avg{\beta_T}$ or
the kinetic freeze-out temperature $T_{\mathrm{kin}}$ was changed with the
yield of each species fixed, and $\lamNO$ was obtained with the fit of
Sec.~\ref{sec:rescaling}. The value 0.9692 of the 3--4\% class corresponds
to $\Delta\avg{\beta_T}=-0.008$ to $-0.009$ at fixed $T_{\mathrm{kin}}$,
or to $\Delta T_{\mathrm{kin}}=-4.7$ to $-5.2$~MeV at fixed
$\avg{\beta_T}$. The two changes differ in shape: with the uncertainties
of the 3--4\% class, a pure stretch describes the change of
$T_{\mathrm{kin}}$ with $\chi^2=2.2$--3.9 for 18 degrees of freedom, but
the change of $\avg{\beta_T}$ only with $\chi^2=52$--70, so the data
($\chi^2=5.9$) resemble a change of the freeze-out temperature more than
a change of the flow velocity alone; a combined change of the two
parameters is not excluded. In both cases pions are stretched most and
protons least, more strongly so for a change of the temperature.

\subsection{Multiplicity ratio at equal mean transverse momentum}
\label{sec:ruler}
\begin{figure}[!htb]
\centering
\includegraphics[width=\columnwidth]{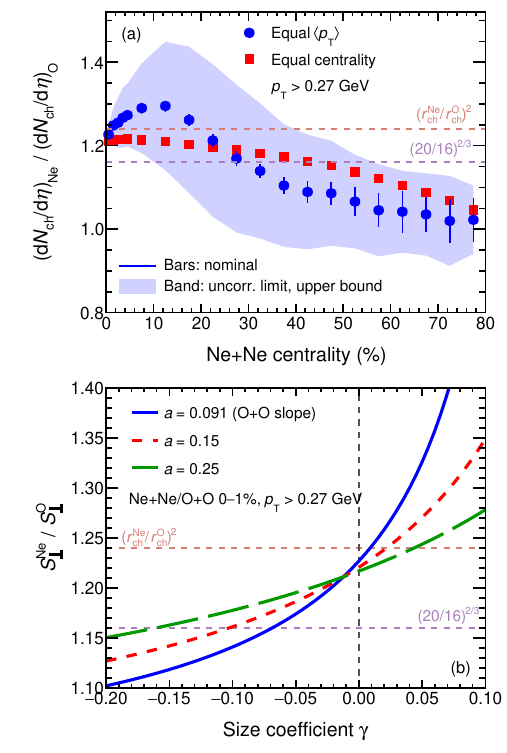}
\caption{(a) Ne+Ne/O+O multiplicity ratio at equal $\avg{\pT}$,
$k$, and at equal centrality, for $\pT>0.27$~GeV, versus the Ne+Ne
centrality; bars: nominal uncertainty; band: upper bound in the
uncorrelated limit. (b) Ratio of the effective transverse areas in the
0--1\% class, Eq.~\eqref{eq:area}, versus the size coefficient $\gamma$
for three values of the density coefficient $a$ ($a=0.091$: O+O slope
between the 0--1\% and 1--2\% classes; $a=0.15$: intermediate; $a=0.25$: close to the ultra-central
Pb+Pb slope~\cite{CMS:2024sgx}); the relative uncertainty from the published
ratios is at most 0.6\% in the range shown, and the vertical line marks
$\gamma=0$. Dashed horizontal lines: $(20/16)^{2/3}$ and the squared ratio
of the rms charge radii of $^{20}$Ne and $^{16}$O~\cite{Angeli:2013epw}.}
\label{fig:area}
\end{figure}

The comparison can also be inverted: instead of $\avg{\pT}$ at equal
multiplicity, the multiplicities at which the two systems have the same
$\avg{\pT}$ are determined. Figure~\ref{fig:area}(a) shows the factor
\begin{equation}
\begin{gathered}
k=\frac{\dNdeta|_{\mathrm{Ne}}(c)}{\dNdeta|^{*}_{\mathrm{O}}},\\[2pt]
\avg{\pT}_{\mathrm{O}}\bigl(\dNdeta|^{*}_{\mathrm{O}}\bigr)=\avg{\pT}_{\mathrm{Ne}}(c),
\end{gathered}
\label{eq:k}
\end{equation}
by which the multiplicity of a Ne+Ne class $c$ exceeds the multiplicity
$\dNdeta|^{*}_{\mathrm{O}}$ at which O+O collisions have the same
$\avg{\pT}$, for $\pT>0.27$~GeV. The O+O $\avg{\pT}$ is interpolated
linearly in $\ln\dNdeta$, and the Ne+Ne values are obtained from the O+O
values of the same class with the published equal-centrality ratios
$N=\dNdeta|_{\mathrm{Ne}}(c)/\dNdeta|_{\mathrm{O}}(c)$ and
$M=\avg{\pT}_{\mathrm{Ne}}(c)/\avg{\pT}_{\mathrm{O}}(c)$. If the O+O
$\avg{\pT}$ rises with a logarithmic slope $a$ near the crossing,
Eq.~\eqref{eq:k} gives $k=N M^{-1/a}$. Unlike $\lamNO$, $k$ can also be
determined for the Ne+Ne 0--3\% classes, because their $\avg{\pT}$ lies
within the range of the O+O values. For the 0--1\% class,
$k=1.227\pm0.005$. The nominal uncertainty is built with the same
correlation assumptions as for the spectra; in the uncorrelated limit an
upper bound of $\pm0.020$ is obtained by shifting the Ne+Ne $\avg{\pT}$
relative to all O+O values by $\sqrt{2}$ times the O+O systematic
uncertainty. The multiplicity ratio at equal centrality is 1.211 in this
class, and since the $\avg{\pT}$ values at equal centrality differ by
only 0.12\% ($M=0.99882$), $k$ is set mainly by $N$. Between 1 and 15\%
centrality $k$ is 1.25--1.30 and then decreases towards unity. A ratio of
transverse areas is not expected to behave in this way: outside the most
central class the small slope $a$ amplifies the small deviation of $M$
from unity, so only the 0--1\% value is discussed below.

The picture of Sec.~\ref{sec:interp} turns the factor $k$ into an
estimate of the area ratio. The mean transverse momentum is
parametrised as
\begin{equation}
\ln\avg{\pT}=a\ln\frac{\dNdeta}{S_\perp}+\frac{\gamma}{2}\ln S_\perp+\mathrm{const},
\label{eq:param}
\end{equation}
where $S_\perp$ is the effective transverse area, the density coefficient $a$ describes the
dependence on the transverse density, and the size coefficient $\gamma$ the dependence on the
size at fixed density: with $S_\perp\propto R^2$,
$\avg{\pT}\propto R^{\gamma}$ at fixed density. The coefficient $a$ is
empirical and is not identified with the squared speed of sound $c_s^2$:
the O+O slope is measured between 1\%-wide centrality classes, in which
the geometry is only approximately fixed, and even at fixed geometry the
identification is not guaranteed~\cite{Gavassino:2025bts}.
For a Ne+Ne and an O+O
class of the same centrality, with $N$ and $M$ as above,
Eq.~\eqref{eq:param} gives the ratio of the effective areas,
\begin{equation}
\frac{S_\perp^{\mathrm{Ne}}}{S_\perp^{\mathrm{O}}}=
\exp\!\left(\frac{a\ln N-\ln M}{a-\gamma/2}\right).
\label{eq:area}
\end{equation}
For $\gamma=0$, $\avg{\pT}$ depends on the density only, equal $\avg{\pT}$
means equal density, and the area ratio is the multiplicity ratio
at equal $\avg{\pT}$: with $a=0.091$, the O+O slope between the two
most central classes, Eq.~\eqref{eq:area} gives 1.227 for the 0--1\%
class, the value of $k$. Figure~\ref{fig:area}(b) shows Eq.~\eqref{eq:area}
for this class ($N=1.2112$, $M=0.99882$). For $\gamma=0$ the area ratio is
1.217--1.227 for slopes $a$ between 0.091 and 0.25, the latter close to
the slope $0.241\pm0.016$ measured in ultra-central Pb+Pb
collisions~\cite{CMS:2024sgx}, larger than
$(20/16)^{2/3}=1.160$, the value for radii proportional to $A^{1/3}$,
and close to 1.240, the squared ratio of the measured rms charge radii
of $^{20}$Ne (3.0055~fm) and $^{16}$O (2.6991~fm)~\cite{Angeli:2013epw}.
A positive $\gamma$ increases the area ratio, and Eq.~\eqref{eq:area}
diverges at $\gamma=2a$, so the result is very sensitive to $\gamma$; the
value 1.160 would require $\gamma$ between $-0.07$ and $-0.16$, and
$\gamma$ is not determined here. The agreement with the charge radii may
also be partly accidental, since at $\gamma=0$ the area ratio is close to
$N$, which in central collisions is close to the ratio of the numbers of
participating nucleons. The charge radii include the charge distribution
of the proton and are not the radii of the nucleon density, and the
effective area also includes the nucleon width, which brings the expected
ratio closer to unity. In addition, the charge-radius value lies within the
upper bound on the uncertainty of $k$ in the uncorrelated limit, and $S_\perp$
is the effective area of the collisions selected as 0--1\% most central,
which is related to the nuclear radii only through an initial-state
model~\cite{Rybczynski:2019adt,Loizides:2025ule}. The comparison with
the charge radii is therefore only indicative and does not measure
the nuclear size.

\section{Summary and outlook}
\label{sec:summary}
The charged-hadron spectra published by the ATLAS
Collaboration~\cite{ATLAS:2026zgq} for Ne+Ne and O+O collisions at
$\sNN=5.36$~TeV have been compared at the same multiplicity, with the O+O
measurements interpolated to the $\dNdeta$ of each Ne+Ne centrality class
from 3 to 80\%. At the same multiplicity the Ne+Ne spectrum is softer.
The ratio of the published $\avg{\pT}$ values is 0.9822 for Ne+Ne 3--4\%
and 0.9918 for Ne+Ne 10--15\%, $6.6\sigma$ and $3.3\sigma$ below unity
even if the O+O systematic uncertainties are uncorrelated between
centrality classes. At $0.4<\pT<2$~GeV the difference is described by a
stretch factor $\lamNO=0.9692\pm0.0032$ for Ne+Ne 3--4\%, $9.5\sigma$
below unity if the common part of the O+O systematic uncertainties cancels
between classes and $3.3\sigma$ ($\pm0.0094$) if it does not; it rises
towards unity with decreasing multiplicity and is consistent with unity
from about 35\% centrality onwards. The forward-pseudorapidity spectra give the
same stretch factors within 0.12\%, and the Xe+Xe/Pb+Pb control shows that
the method itself does not produce a softening as large as the observed one.

Since the spectral shapes at equal centrality are nearly the same, in
central collisions the spectral shape follows the centrality class more
closely than the multiplicity. The ordering is the one expected if a larger
transverse area at the same multiplicity gives a softer spectrum, as in
hydrodynamic and initial-state saturation pictures, which the data do not
distinguish. In a blast-wave description, the stretch of the 3--4\% class
corresponds to a decrease of the kinetic freeze-out temperature by about
5~MeV or of the mean transverse flow velocity by 0.008--0.009, and its
shape is closer to that of a temperature change. In the 0--1\% class, beyond the O+O multiplicity range,
Ne+Ne collisions reach the same $\avg{\pT}$ as O+O collisions at a 23\%
larger multiplicity ($k=1.227\pm0.005$).

{\setlength{\parskip}{0pt}%
Calculations of the transverse areas of O+O and Ne+Ne collisions at
matched multiplicity, and hydrodynamic or transport calculations of the
spectra analysed in the same way, are needed to turn the present results
into a constraint on the structure of $^{16}$O and $^{20}$Ne. A published
breakdown of the O+O systematic uncertainties by source would decide
between the two significance estimates given here. Identified-particle
spectra in fine centrality classes would allow the same comparison for
pions, kaons and protons separately, and thus distinguish a change of the
flow velocity from one of the freeze-out temperature, which predict
different mass orderings.\par}

\begin{acknowledgments}
The author thanks Dr.~Somadutta Bhatta for helpful discussions on the systematic uncertainties of the ATLAS data. This work is supported by OSHEC, Department of Higher Education, Government of Odisha, Index No.~23EM/PH/124, under MRIP 2023.
\end{acknowledgments}

\section*{Data availability}
All inputs are public in HEPData: records 181329.v1~\cite{ATLAS:2026zgq}, 85727.v3~\cite{ALICE:2018hza}, 86210.v1~\cite{ALICE:2018vuu}, 153671.v1~\cite{ATLAS:2024jvf}, 104923.v1~\cite{ALICE:2019hno} and 146016.v1~\cite{CMS:2024sgx}. The analysis code is available from the author upon reasonable request.

\appendix
\section{Stretch factors and multiplicity matching}
\label{app:matching}
\begin{table*}[!t]
\centering\footnotesize
\caption{Stretch factor $\lamNO$ of the Ne+Ne spectrum relative to the O+O spectrum at the same $\dNdeta$, Eq.~\eqref{eq:resc}. For $|\eta|<0.5$: total nominal uncertainty $\sigma_{\mathrm{nom}}$, total uncertainty in the uncorrelated limit $\sigma_{\mathrm{unc}}$, and the deviation from unity in units of each; for $1.5<|\eta|<2.25$: nominal uncertainty. $\dNdeta$ is that of the Ne+Ne class at $|\eta|<0.5$, extrapolated to $\pT=0$ (Table~\ref{tab:matching}).}
\label{tab:lambda}
\renewcommand{\arraystretch}{1.0}
\begin{tabular*}{\textwidth}{@{\extracolsep{\fill}}ccccccccc}
\toprule
 & & \multicolumn{5}{c}{$|\eta|<0.5$} & \multicolumn{2}{c}{$1.5<|\eta|<2.25$}\\
\cmidrule(lr){3-7}\cmidrule(lr){8-9}
Ne+Ne class & $\dNdeta$ & $\lamNO$ & $\sigma_{\mathrm{nom}}$ & $\sigma_{\mathrm{unc}}$ & $(1-\lamNO)/\sigma_{\mathrm{nom}}$ & $(1-\lamNO)/\sigma_{\mathrm{unc}}$ & $\lamNO$ & $\sigma_{\mathrm{nom}}$\\
\midrule
3--4\% & 145.0 & 0.9692 & 0.0032 & 0.0094 & 9.5 & 3.3 & 0.9692 & 0.0042\\
4--5\% & 139.3 & 0.9718 & 0.0030 & 0.0093 & 9.5 & 3.0 & 0.9712 & 0.0035\\
5--10\% & 125.7 & 0.9785 & 0.0031 & 0.0091 & 6.9 & 2.4 & 0.9774 & 0.0030\\
10--15\% & 107.5 & 0.9852 & 0.0042 & 0.0096 & 3.5 & 1.5 & 0.9854 & 0.0026\\
15--20\% & 92.7 & 0.9903 & 0.0024 & 0.0089 & 4.1 & 1.1 & 0.9900 & 0.0025\\
20--25\% & 80.0 & 0.9931 & 0.0027 & 0.0090 & 2.6 & 0.8 & 0.9932 & 0.0027\\
25--30\% & 68.9 & 0.9948 & 0.0029 & 0.0088 & 1.8 & 0.6 & 0.9949 & 0.0029\\
30--35\% & 59.2 & 0.9957 & 0.0033 & 0.0090 & 1.3 & 0.5 & 0.9956 & 0.0030\\
35--40\% & 50.5 & 0.9958 & 0.0040 & 0.0092 & 1.0 & 0.5 & 0.9966 & 0.0037\\
40--45\% & 43.0 & 0.9963 & 0.0052 & 0.0096 & 0.7 & 0.4 & 0.9966 & 0.0051\\
45--50\% & 36.3 & 0.9958 & 0.0058 & 0.0099 & 0.7 & 0.4 & 0.9966 & 0.0055\\
50--55\% & 30.5 & 0.9970 & 0.0063 & 0.0098 & 0.5 & 0.3 & 0.9973 & 0.0063\\
55--60\% & 25.5 & 0.9973 & 0.0077 & 0.0102 & 0.3 & 0.3 & 0.9970 & 0.0078\\
60--65\% & 21.2 & 0.9974 & 0.0084 & 0.0104 & 0.3 & 0.2 & 0.9972 & 0.0089\\
65--70\% & 17.5 & 0.9967 & 0.0098 & 0.0105 & 0.3 & 0.3 & 0.9979 & 0.0102\\
70--75\% & 14.2 & 0.9974 & 0.0116 & 0.0118 & 0.2 & 0.2 & 0.9977 & 0.0113\\
75--80\% & 11.4 & 0.9973 & 0.0155 & 0.0156 & 0.2 & 0.2 & 0.9980 & 0.0149\\
\bottomrule
\end{tabular*}
\end{table*}

Table~\ref{tab:lambda} lists the stretch factor $\lamNO$ of all seventeen compared Ne+Ne classes, at mid and forward pseudorapidity, with its nominal uncertainty and, at mid-rapidity, the uncertainty in the uncorrelated limit. Table~\ref{tab:matching} gives, for each centrality class, the multiplicities of the two systems, the published ratio $r_{\mathrm{Ne/O}}$, and the two O+O classes and the weight used in Eq.~\eqref{eq:interp} to obtain the O+O spectrum at the multiplicity of the Ne+Ne class. Since $r_{\mathrm{Ne/O}}>1$, each Ne+Ne class is matched by more central O+O classes or, from 40\% centrality onwards, by the O+O class of the same centrality and the next more central one. The ratio is 1.21--1.22 in the classes up to 10\% centrality and decreases to 1.05 in the 75--80\% class, so towards peripheral collisions the matched O+O classes approach the Ne+Ne class of the same centrality.

\begin{table}[!htb]
\centering\footnotesize
\caption{Multiplicity matching. For each centrality class: $\dNdeta$ ($|\eta|<0.5$, extrapolated to $\pT=0$) in O+O and Ne+Ne, the latter from Eq.~\eqref{eq:match}; the published ratio $r_{\mathrm{Ne/O}}$; the O+O classes 1 and 2 (centrality in \%) that bracket the Ne+Ne multiplicity, and the weight $w_1=1-w_2$ of class 1 in Eq.~\eqref{eq:interp}. Dashes: the Ne+Ne multiplicity lies above that of the most central O+O class, and the class is not used.}
\label{tab:matching}
\begin{tabular*}{\columnwidth}{@{\extracolsep{\fill}}cccccc}
\toprule
Centrality & \multicolumn{2}{c}{$\dNdeta$} & $r_{\mathrm{Ne/O}}$ & O+O classes & $w_1$\\
 & O+O & Ne+Ne & & 1, 2 & \\
\midrule
0--1\% & 148.4 & 179.6 & 1.2101 & --- & ---\\
1--2\% & 133.0 & 161.3 & 1.2132 & --- & ---\\
2--3\% & 125.2 & 152.1 & 1.2142 & --- & ---\\
3--4\% & 119.5 & 145.0 & 1.2134 & 0--1, 1--2 & 0.79\\
4--5\% & 114.7 & 139.3 & 1.2151 & 0--1, 1--2 & 0.43\\
5--10\% & 103.7 & 125.7 & 1.2124 & 1--2, 2--3 & 0.06\\
10--15\% & 89.0 & 107.5 & 1.2079 & 4--5, 5--10 & 0.36\\
15--20\% & 77.2 & 92.7 & 1.2007 & 5--10, 10--15 & 0.26\\
20--25\% & 67.0 & 80.0 & 1.1934 & 10--15, 15--20 & 0.25\\
25--30\% & 58.1 & 68.9 & 1.1864 & 15--20, 20--25 & 0.19\\
30--35\% & 50.2 & 59.2 & 1.1792 & 20--25, 25--30 & 0.13\\
35--40\% & 43.2 & 50.5 & 1.1700 & 25--30, 30--35 & 0.05\\
40--45\% & 37.0 & 43.0 & 1.1598 & 35--40, 40--45 & 0.96\\
45--50\% & 31.6 & 36.3 & 1.1490 & 40--45, 45--50 & 0.88\\
50--55\% & 26.9 & 30.5 & 1.1339 & 45--50, 50--55 & 0.78\\
55--60\% & 22.8 & 25.5 & 1.1192 & 50--55, 55--60 & 0.68\\
60--65\% & 19.2 & 21.2 & 1.1027 & 55--60, 60--65 & 0.57\\
65--70\% & 16.1 & 17.5 & 1.0846 & 60--65, 65--70 & 0.46\\
70--75\% & 13.3 & 14.2 & 1.0682 & 65--70, 70--75 & 0.35\\
75--80\% & 10.9 & 11.4 & 1.0456 & 70--75, 75--80 & 0.22\\
\bottomrule
\end{tabular*}
\end{table}

\bibliography{NeO_softeningNotes}

\end{document}